\documentclass[aip,pop,preprint]{revtex4-1}
\usepackage[utf8]{inputenc}
\usepackage{amsmath,amssymb,siunitx}
\usepackage{graphicx,float,dcolumn,bm}
\usepackage{hyperref,comment}
\usepackage{natbib}
\usepackage{caption}
\usepackage{subcaption}
\usepackage{adjustbox}
\usepackage{booktabs}
\begin{document}

\newcommand{\victor}[1]{{\textcolor{blue}{#1}}}

\title[Shock Wave Formation in Radiative Plasmas]{Shock Wave Formation in Radiative Plasmas}

\author{F. Garcia-Rubio}
\affiliation{Laboratory for Laser Energetics, Rochester, New York 14623, USA}
\affiliation{University of Rochester, Rochester, New York 14627, USA}
\affiliation{Currently at Pacific Fusion Corporation, Fremont, California 94538, USA}
\author{V. Tranchant}
\affiliation{University of Rochester, Rochester, New York 14627, USA}
\author{E. C. Hansen}
\affiliation{University of Rochester, Rochester, New York 14627, USA}
\author{R. Tabassum}
\affiliation{Preston University Islamabad, Pakistan}
\author{A. Reyes}
\affiliation{University of Rochester, Rochester, New York 14627, USA}
\author{H. U. Rahman}
\affiliation{Magneto-Inertial Fusion Technology Inc., Tustin, California 92780, USA}
\author{P. Ney}
\affiliation{Magneto-Inertial Fusion Technology Inc., Tustin, California 92780, USA}
\author{E. Ruskov}
\affiliation{Magneto-Inertial Fusion Technology Inc., Tustin, California 92780, USA}
\author{P. Tzeferacos}
\affiliation{University of Rochester, Rochester, New York 14627, USA}
\affiliation{Laboratory for Laser Energetics, Rochester, New York 14623, USA}

\date{\today}

\begin{abstract}

The temporal evolution of weak shocks in radiative media is theoretically investigated in this work. 
The structure of radiative shocks has traditionally been studied in a stationary framework. 
Their systematic classification is complex because layers of optically thin and thick regions alternate to form a radiatively-driven precursor and a temperature-relaxation layer, between which the hydrodynamic shock is embedded. 
In this work, we analyze the formation of weak shocks when two radiative plasmas with different pressures are put in contact.
Applying a reductive perturbative method yields a Burgers-type equation that governs the temporal evolution of the perturbed variables including the radiation field. 
The conditions upon which optically thin and thick solutions exist have been derived and expressed as a function of the shock strength and Boltzmann number.
Below a certain Boltzmann number threshold, weak shocks always become optically thick asymptotically in time, while thin solutions appear as transitory structures.
The existence of an optically thin regime is related to the presence of an overdense layer in the compressed material. 
Scaling laws for the characteristic formation time and shock width are provided for each regime. 
The theoretical analysis is supported by \textit{FLASH} simulations, and a comprehensive test-case has been designed to benchmark radiative hydrodynamic codes.  

\end{abstract}

\maketitle

\section{\label{sec:intro}Introduction}

The Z-pinch concept typically relies on the use of a pulsed-power generator to magnetically compress a target load onto the axis in which the current flows.\cite{haines2011review,giuliani2015review}
The compression of the material is mediated by one or successive shock waves propagating in a collisional plasma, followed by weaker compressional waves. 
The  structure of such waves is determined by the nature of the dissipative mechanisms in play.
Among the different applications to fusion energy,\cite{slutz2010pulsed,slutz2012high} the staged Z-pinch approach\cite{Rahman2009Plasma,Wessel2015,Wessel2016,ruskov2021staged} seeks to exploit the radiative nature of shocks that develop in high-atomic-number liners.
Understanding the structure of said compressional waves becomes important for the design of such configurations.

Radiation and other dissipation mechanisms compete to conform the shock structure in an intricate manner.
In unmagnetized plasmas, composed of ions and electrons of disparate masses $m_i$ and $m_e$, respectively, the thermal conductivity of the electrons is larger than that of the ions by a factor $\left(m_i/m_e\right)^{1/2}$. \cite{Braginskii1965}
However, since electrons carry little bulk kinetic energy due to their small mass, most of the shock heating comes from the kinetic energy of the incoming ion flow. 
The plasma density jump takes place in a length scale of the order of the ion mean free path $\lambda$, where ion heat conduction and viscosity operate similarly.\cite{mihalas2013foundations} 
The electrons, on their part, equilibrate with the ions in a wider layer of thickness $\left(m_i/m_e\right)^{1/2}\lambda$ in which the compression shock is embedded.\cite{shafranov}
The structure of radiative shock fronts is somewhat more complex since the mean free path of the photons $l_p$ generally introduces a larger scale where the radiation is emitted and absorbed. \cite{drake2010high} 
Ions and electrons are therefore compressed in a thin hydrodynamic shock embedded between a precursor upstream and a relaxation region downstream. 

The difficulty in modeling radiative shock fronts stems from the fact that their characteristic length $L$ depends on the regime in which matter and radiation interact, while, reciprocally, this regime is dictated by how $L$ compares to the mean free path of the photons; i.e., its optical depth $\tau = L/l_p$. 
In optically thick regions, $\tau \gg 1$, matter and radiation are in local thermal equilibrium, and the radiation effect on the hydrodynamics possesses the structure of radiation heat conduction.
In optically thin regions, $\tau \ll 1$, radiation energy density is homogeneous and the radiant heat exchange acts as a cooling function for the flow. 


\begin{figure}[H]
    \centering
    \includegraphics[width=1\linewidth]{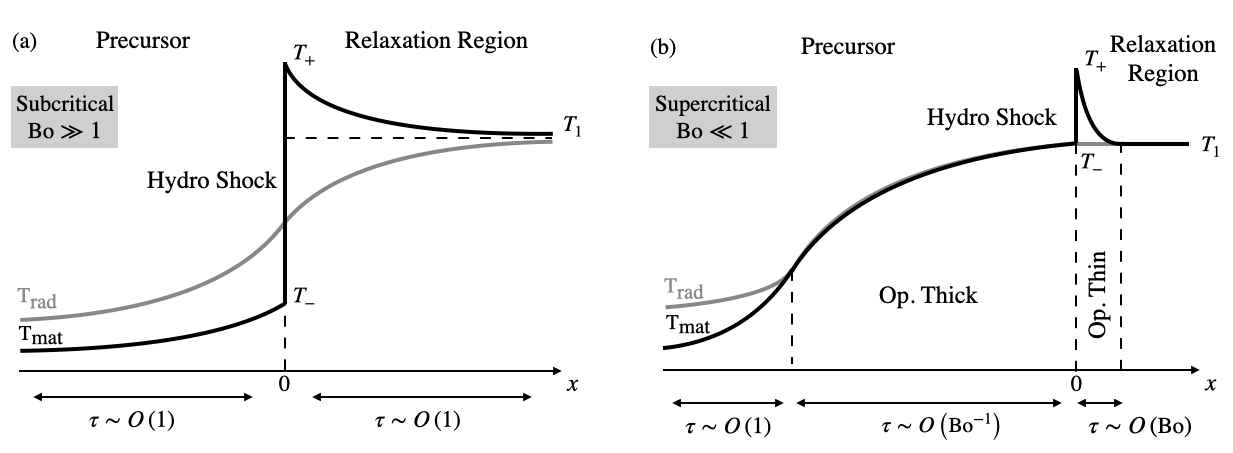}
    \caption{Schematic of (a) subcritical and (b) supercritical shock fronts. Flow is moving from left to right. Matter temperature is in black and radiation temperature is in gray. The temperature jump $T_+-T_-$ satisfies Rankine-Hugoniot conditions. The Boltzmann number Bo is based on the flow conditions downstream. In the subcritical shock, one can estimate $O\left( T_- / T_1 \right)\sim O\left(\text{Bo}^{-1}\right)$. The supercritical shock sketched is characterized by Bo $\ll 1$, resulting in $T_-\approx T_1$.} 
    \label{fig:OTRsketch}
\end{figure}

Traditionally, radiative shock studies have considered a stationary flow and and an optically thick medium far upstream and downstream. \cite{drake2007theory,lowrie2008radiative,masser2011shock}
In Zel'dovich and Raizer,\cite{zel2002physics} the shock-front phenomenology is divided into subcritical and supercritical cases, shown in Fig. \ref{fig:OTRsketch}. 
Although not explicitly stated, this classification is based on the Boltzmann number (Bo) of the compressed gas (downstream), which accounts for the ratio between enthalpy and radiative fluxes, $\text{Bo}\sim p c_s / \sigma T^4$.
In the former case, matter and radiation are out of equilibrium across the shock front, its width being of the order of the photon mean free path, $\tau \sim O(1)$. 
The supercritical shock presents a richer structure, as an optically thick layer follows the nonequilibrium heating region leading the precursor, Fig. \ref{fig:OTRsketch}(b). 
Beyond the hydrodynamic shock, the heated matter cools in an optically thin layer (Zel'dovich spike) until matter and radiation come to equilibrium forming the relaxation region.
In spite of this complexity, the density profile is monotonic and matter is not compressed beyond what Rankine-Hugoniot conditions dictate. 

Several efforts to provide a systematic classification of radiative shocks can be found in the literature. 
\citet{bouquet2000analytical} considered optically thick shocks retaining radiation pressure effects and based their analysis on the Mach number and radiation-to-thermal pressure ratio $\alpha$. 
Later, \citet{michaut2009classification} proposed to simultaneously use three dimensionless parameters related to Bo, $\tau$ and $\alpha$.
\citet{falize2009analytical} identified five different cases when deriving scaling laws for optically thin shocks. 
In a similar spirit, Lie groups theory has been used to derive scaling laws for radiating fluids encompassing optically thin and thick regimes.\cite{falize2009scaling,tranchant2022new}

In this paper, we complement the current understanding of radiative shocks by investigating  the temporal evolution of weak shocks in radiative media. 
We specifically analyse the shock formation process after two radiative plasmas with a small pressure imbalance are put in contact.
In the weak limit, the thickness of a shock scales inversely proportional to its strength. \cite{zel2002physics}
This represents a significant simplification, as different species have time to equilibrate across the shock. 
We have identified the regimes where radiative effects dominate over thermal conduction, and assessed the conditions for which the shock develops an optically thin or thick structure.
These conditions are expressed in terms of the Boltzmann number of the medium and the shock strength.
Our analysis is strongly influenced by the seminal work of Hu\cite{hu1966structure,hu1972collisional} and \citet{grad1967unified}, where the formation of weak plasma shocks in the presence of a magnetic field including all possible dissipation mechanisms was systematically investigated. 
Similar analysis has been performed thereafter in various scenarios related to plasma physics,\cite{bujarbarua1980magnetosonic,boldyrev1998nonlinear,saeed2010nonlinear,hussain2011korteweg} and magnetic flux transport in staged Z-pinch implosions. \cite{narkis2017investigation}
To the authors' knowledge, however, this is the first time that radiative effects are considered. 
We therefore disregard the presence of magnetic fields and focus on the interaction between matter and radiation.

This paper is organized as follows. In Sec. \ref{sec:governing equations}, the governing equations are presented.
In Sec. \ref{sec:weak shocks}, the perturbation method is applied and the formation of weak shocks is discussed. 
The theoretical analysis is compared to numerical simulations using the radiation-magnetohydrodynamic code \emph{FLASH} in Sec. \ref{Sec:numerical}.
Finally, in Sec. \ref{Sec: conclusions}, conclusions are drawn and the application of the theory to the design of Z-pinch implosions is discussed. 

\section{\label{sec:governing equations}Governing equations}

We consider a hydrodynamic description of an unmagnetized plasma coupled to radiation.
We assume a quasi-neutral plasma with same ion and electron temperatures, and neglect the radiation pressure terms. 
In order to analyze the physics of interest, we propose Cartesian coordinates and assume that all quantities depend exclusively on time $t$ and (streamwise) spatial coordinate $x$. 
The only nonzero fluid velocity component is in the streamwise direction $\textbf{v}=u\left(t,x\right)\textbf{e}_{x}$. 

With these assumptions, the equations governing the plasma mass density $\rho$, temperature $T$, and velocity $u$ correspond to continuity, momentum, and internal energy conservation, reading:

\begin{equation}
    \dfrac{\partial\rho}{\partial t} + 
    \dfrac{\partial}{\partial x}\left(\rho u\right) = 0,
    \label{eq:contDim}
\end{equation}

\begin{equation}
    \rho\dfrac{\partial u}{\partial t} + 
    \rho u\dfrac{\partial u}{\partial x} = 
    - \dfrac{\partial p}{\partial x},
    \label{eq:momentumDim}
\end{equation}

\begin{equation}
    \dfrac{\left( Z+1 \right)\rho}{\left(\gamma - 1\right) m_i}\left( \dfrac{\partial T}{\partial t} + u \dfrac{\partial T}{\partial x} \right)
     = 
     - p \dfrac{\partial u}{\partial x} +
    \dfrac{\partial}{\partial x}\left(\chi\dfrac{\partial T}{\partial x}\right) -
    \phi _{eR}.
    \label{eq:energyDim}
\end{equation}
Here, $Z$ refers to the ionic charge, $\gamma$ is the adiabatic index,  $m_i$ is the ion mass, and $\chi$ the electron thermal conductivity.
For the latter, we follow the formulation and notations used in \citet{Braginskii1965}. 
The thermal pressure $p$ satisfies the equation of state $p = \left( Z+1 \right)\rho T /m_i$, and the temperature $T$ is expressed in energy units. 
The term $\phi _{eR}$ refers to the variation of electron internal energy due to the total emission and absorption of radiation, that is,

\begin{equation}
    \phi _{eR} = c K_P \left( U_P - U_R \right). 
    \label{eq:Coupling Term}
\end{equation}
Here, $c$ is the speed of light, $K_P$ refers to the Planck mean opacity, $U_P$ is the planckian radiation energy density, $U_P=4\sigma T^4/c$, with $\sigma$ the Stefan-Boltzmann constant, and $U_R$ stands for the radiation energy density. 
For convenience, we express it in terms of the radiation temperature $T_R$ as $U_R=4\sigma T_R^4/c$.

In order to close the system of equations, the radiative transfer equation is used under the assumptions of a steady state, and isotropic and elastic scattering. 
It relates the radiation energy flux $S$ to the emission and absorption of radiation,

\begin{equation}
    \dfrac{\partial S}{\partial x} = c K_P \left( U_P - U_R \right).
    \label{eq:radiationDim}
\end{equation}
Within the framework of the diffusion approximation assumed in this analysis, we can express the radiation energy flux as 

\begin{equation}
    S = - \dfrac{c}{3 K_R} \dfrac{\partial U_R}{\partial x},
    \label{eq:Rad Energy Flux}
\end{equation}
where we have taken a constant Eddington factor equal to $1/3$, corresponding to an isotropic angular distribution of the radiation.
The coefficient $K_R$ stands for the Rosseland mean opacity. 
For simplicity, we will assume both $K_P$ and $K_R$ to be equal to the inverse of the photon mean free path, $l_p$, and denote them as $K=1/l_p$ hereinafter. 

\subsection*{\label{subsec:normalization}Normalization}

In order to derive a governing equation for weak nonlinear waves, we start by normalizing Eqs. \eqref{eq:contDim} -- \eqref{eq:energyDim}  and \eqref{eq:radiationDim} with a characteristic length $L$, density $\rho _0$, temperature $T_0$, and speed of sound $c_0 = \sqrt{\gamma\left(Z+1\right)T_0 / m_i}$. 
Pressure is normalized with $p_0 \equiv \rho _0 c_0 ^2/\gamma$. 
Next, we stretch the coordinates as 

\begin{equation}
x^{\prime}=\epsilon\left(\dfrac{x}{L}+v_{0}\dfrac{c_0 t}{L}\right),\quad t^{\prime}=\epsilon^{2}\dfrac{c_0 t}{L},\label{eq:stretching}
\end{equation}
where the small parameter $\epsilon\ll1$ characterizes the strength of the nonlinearity, and $v_{0}$ represents its phase velocity, eigenvalue to be determined during the resolution.
The scaling proposed for the dimensionless time $t^{\prime}$ is such that unsteady, dissipative, and nonlinear convective effects are equally important.\cite{hu1972collisional}

Keeping, for simplicity, the same name for the dimensionless quantities as their non-normalized counterpart, the dimensionless governing equations become

\begin{equation}
    \epsilon\dfrac{\partial\rho}{\partial t^{\prime}} +
    \left(u+v_0\right) \dfrac{\partial\rho}{\partial x^{\prime}}+
    \rho \dfrac{\partial u}{\partial x^{\prime}} = 0,
    \label{eq:contNorm}
\end{equation}

\begin{equation}
    \epsilon \rho \dfrac{\partial u}{\partial t^{\prime}} + 
    \rho \left(u+v_0\right) \dfrac{\partial u}{\partial x^{\prime}} +
    \dfrac{1}{\gamma}\dfrac{\partial}{\partial x^{\prime}} \left(\rho T\right) = 0,
    \label{eq:momentumNorm}
\end{equation}

\begin{equation}
    \dfrac{\epsilon}{\gamma-1} \rho \dfrac{\partial T}{\partial t^{\prime}} +
    \dfrac{1}{\gamma -1}\left(u+v_0\right)\rho\dfrac{\partial T}{\partial x^{\prime}} +
    \rho T \dfrac{\partial u}{\partial x^{\prime}} =
    \epsilon \dfrac{\partial}{\partial x^{\prime}}\left(\kappa\dfrac{\partial T}{\partial x^{\prime}}\right) -
    \dfrac{1}{\epsilon}\dfrac{\tau}{\text{Bo}}\left(T^4-T_R^4\right),
    \label{eq:energyNorm}
\end{equation}

\begin{equation}
    \epsilon\dfrac{\partial}{\partial x^{\prime}}\left(\dfrac{1}{3\tau}\dfrac{\partial T_R^4}{\partial x^{\prime}}\right) =
    -\dfrac{\tau}{\epsilon}\left(T^4-T_R^4\right),
    \label{eq:radiationNorm}
\end{equation}
where we have introduced the optical thickness $\tau = KL$, and the normalized conductivity $\kappa = \chi m_i/\left(Z+1\right)\rho_0 c_0 L$. 
They can be expressed in terms of the photon mean free path, $l_p$, and a characteristic length related to electron thermal conductivity, $l_c$,  as $\tau = L/l_p$ and $\kappa = l_c / L$, respectively. 
The Boltzmann number Bo accounts for the ratio between enthalpy and radiative fluxes

\begin{equation}
    \text{Bo} = \dfrac{p_0c_0}{4\sigma T_0^4}.
    \label{eq:Boltzmann}
\end{equation}

For practical purposes, numerical expressions of Bo, $l_c$ and $l_p$ are provided below. 
The latter has been taken to be equal to the Planck mean free path for the case of a fully ionized gas with absorption only by bremsstrahlung [Eq. (5.23) in \citet{zel2002physics}] :

\begin{equation}
    \text{Bo} \approx 2.30 \gamma ^{1/2} \left(\dfrac{Z+1}{A}\right)^{3/2}\dfrac{\rho _0}{\text{1 g/cm}^3}\left(\dfrac{T_0}{\text{100 eV}} \right)^{-5/2},
    \label{eq:BoltzmannNum}
\end{equation}

\begin{equation}
     \dfrac{l_c}{\text{1 cm}}\approx 1.03 \times 10^{-6} \dfrac{\gamma _0 A}{Z\left(Z+1\right)}\dfrac{10}{\lambda}\sqrt{\dfrac{A}{\gamma\left(Z+1\right)}}\left(\dfrac{\rho _0}{\text{1 g/cm}^3}\right)^{-1}\left(\dfrac{T_0}{\text{100 eV}} \right)^{2},
    \label{eq:lcNum}
\end{equation}

\begin{equation}
   \dfrac{l_p}{\text{1 cm}}\approx 7.35\times 10^{-4}\dfrac{A^2}{Z^3}\left(\dfrac{\rho _0}{\text{1 g/cm}^3}\right)^{-2}\left(\dfrac{T_0}{\text{100 eV}} \right)^{7/2}.
    \label{eq:lpNum}
\end{equation}
Here, $A$ refers to the mass number, $\lambda$ stands for the Coulomb logarithm, and $\gamma _0$ is a coefficient given by \citet{Braginskii1965} that depends exclusively on $Z$. It ranges from 3.1616 for $Z=1$ to 12.471 for $Z \rightarrow \infty$.

\section{\label{sec:weak shocks}Perturbation analysis for weak shocks in radiative plasmas}

We study solutions of the system of Eqs. \eqref{eq:contNorm} -- \eqref{eq:radiationNorm} driven by a downstream overpressure.  
Such solutions develop a shock-type structure of thickness $\sim L/\epsilon$ in a time scale $\sim L/c_0\epsilon ^2$. 
The scaled thickness $L$ must be determined self-consistently and is given by the dissipation mechanisms present in the system: conduction and radiation.
However, we anticipate the existence of three characteristic lengths. 
Thermal conduction introduces $l_c$, but radiation effects result in an optically thin dissipation length $l_{\text{thin}}=l_p\text{Bo}$, and an optically thick dissipation length $l_\text{thick}=l_p/\text{Bo}$. 
The last two regimes do not operate simultaneously, but rather arise depending on how the Boltzmann number scales with the strength of the perturbation $\epsilon$.
The analysis differs in this sense with that of \citet{grad1967unified} and \citet{hu1966structure}, since there is a dimensionless number, the optical thickness $\tau=L/l_p$, whose order of magnitude needs to be estimated \textit{a priori}.
This estimation must be verified after the resolution of the system.

\subsection{\label{subsec:resolution}Resolution method}

We consider at zeroth order a plasma at rest and in thermodynamic equilibrium with the radiative flow. 
We therefore expand the normalized variables as:

\begin{equation}\label{eq:varExpand}
\begin{split}
    \rho\left(t^{\prime},x^{\prime}\right)=1+\sum _{i=1} ^{i=\infty}\epsilon ^i \rho_i\left(t^{\prime},x^{\prime}\right),\quad &u\left(t^{\prime},x^{\prime}\right)=\sum _{i=1} ^{i=\infty}\epsilon ^i u_i\left(t^{\prime},x^{\prime}\right),\\
    T\left(t^{\prime},x^{\prime}\right)=1+\sum _{i=1} ^{i=\infty}\epsilon ^i T_i\left(t^{\prime},x^{\prime}\right),\quad &T_R\left(t^{\prime},x^{\prime}\right)=1+\sum _{i=1} ^{i=\infty}\epsilon ^i T_{R_i}\left(t^{\prime},x^{\prime}\right),
\end{split}
\end{equation}
respectively.
The structure of Eqs. \eqref{eq:contNorm} -- \eqref{eq:radiationNorm} admits for a scaling of the Boltzmann number and optical depth as  

\begin{equation}
    \text{Bo}\sim O\left(\epsilon ^{r/2}\right),\quad \tau \sim O\left(\epsilon ^{s/2}\right)
    \label{eq:BoTau}
\end{equation}
with $r,s$ integers. 
We remark that the first scaling in Eq. \eqref{eq:BoTau} is an input to the problem, as it characterizes the Boltzmann number of the medium in which the wave develops with respect to its strength.
The second scaling is an estimation of the optical depth of the wave, to be verified \textit{a posteriori}. 
Without loss of generality, we can assume the normalized conductivity $\kappa$ in Eq. \eqref{eq:energyNorm} to be of order unity at most. 
Effectively, if conduction dominates, $\kappa \sim O\left(1\right)$ by definition, whereas in the opposite case, radiation would impose a larger length scale and, consequently, $\kappa \ll 1$.
Inserting the ansätze of Eq. \eqref{eq:varExpand} into Eqs. \eqref{eq:contNorm} -- \eqref{eq:radiationNorm} and gathering alike terms in powers of $\epsilon$ allows one to obtain a recursive system of equations. 

We first illustrate the solution procedure neglecting radiation effects. 
The first order in $\epsilon$ of continuity, momentum, and energy equations yields the following system

\begin{equation}
\underbrace{\left(\begin{array}{ccc}
v_{0} & 1 & 0\\
1/\gamma & v_{0} & 1/\gamma\\
0 & 1 & \dfrac{v_{0}}{\gamma-1}
\end{array}\right)}_{A}\cdot\dfrac{\partial}{\partial x^{\prime}}\left\{ \begin{array}{c}
\rho_{1}\\
u_{1}\\
T_{1}
\end{array}\right\} =\left\{ \begin{array}{c}
0\\
0\\
0
\end{array}\right\},
\label{eq:Order1}
\end{equation}
which admits nontrivial solutions when $\text{det}(A)=0$. 
This occurs for $v_0 = 1$, corresponding to weak shocks traveling towards the -$x$ direction at the isentropic speed of sound. 
The perturbed variables are then related to one another as $u_1=-v_0\rho_1$, $T_1=\left(\gamma -1\right)\rho_1$. 
Taking this into account, the order $\epsilon ^2$ gives

\begin{equation}
    \left(
    \begin{array}{ccc}
    v_{0} & 1 & 0\\
    1/\gamma & v_{0} & 1/\gamma\\
    0 & 1 & \dfrac{v_{0}}{\gamma-1}
    \end{array}
    \right)
    \cdot\dfrac{\partial}{\partial x^{\prime}}\left\{ \begin{array}{c}
    \rho_{2}\\
    u_{2}\\
    T_{2}
    \end{array}\right\} =
    \left\{ 
    \begin{array}{c}
    -\dfrac{\partial\rho_{1}}{\partial t^{\prime}}+2v_{0}\rho_{1}\dfrac{\partial\rho_{1}}{\partial x^{\prime}}\\
    v_{0}\dfrac{\partial\rho_{1}}{\partial t^{\prime}}-2\dfrac{\gamma-1}{\gamma}\rho_{1}\dfrac{\partial\rho_{1}}{\partial x^{\prime}}\\
    -\dfrac{\partial\rho_{1}}{\partial t^{\prime}}+\gamma v_{0}\rho_{1}\dfrac{\partial\rho_{1}}{\partial x^{\prime}}+\kappa\left(\gamma-1\right)\dfrac{\partial^{2}\rho_{1}}{\partial x^{\prime 2}}
    \end{array}
    \right\}. 
    \label{eq:Order2}
\end{equation}
By construction, $v_0 = 1$ is an eigenvalue of the left-hand side of Eq. \eqref{eq:Order2}.
Hence, the eigenvector $\mathbf{q}_s=\left\lbrace 1,-\gamma,\gamma - 1 \right\rbrace^T$ satisfying $\left. A\right|_{v_o=1}^T\cdot \mathbf{q}_s=\mathbf{0}$ is a left-annihilator of the system. 
Left-multiplying by $\mathbf{q}_s^T$ and taking $v_0 = 1$ yields a governing equation for the perturbed density,

\begin{equation}
    \dfrac{\partial \rho_1}{\partial t^{\prime}}-\dfrac{\gamma +1}{2}\rho_1\dfrac{\partial \rho_1}{\partial x^{\prime}}=\dfrac{\left(\gamma -1\right)^2}{2\gamma}\kappa \dfrac{\partial ^2 \rho _1}{\partial x^{\prime 2}}.
    \label{eq:Burgers Cond}
\end{equation}
Introducing now the variables 

\begin{equation}
    t^* = t^{\prime}\dfrac{\gamma \left(\gamma+1\right)^2}{4\left(\gamma -1\right)^2\kappa},\quad x^* = x^{\prime}\dfrac{\gamma \left(\gamma+1\right)}{2\left(\gamma -1\right)^2\kappa}
    \label{eq:tstarxstar}
\end{equation}
transforms Eq. \eqref{eq:Burgers Cond} into 

\begin{equation}
    \dfrac{\partial \rho_1}{\partial t^*}-\rho_1\dfrac{\partial \rho_1}{\partial x^*}=\dfrac{1}{2} \dfrac{\partial ^2 \rho _1}{\partial x^{*2}},
    \label{eq:Burgers Cond Norm}
\end{equation}
which corresponds to the well-known Burgers equation. 
It can be seen that the procedure to derive the first-order perturbed field has been formally closed at order $\epsilon ^2$, with higher orders providing corrections as higher-order terms. 

\begin{figure}[H]
    \centering
    \includegraphics[width=0.6\linewidth]{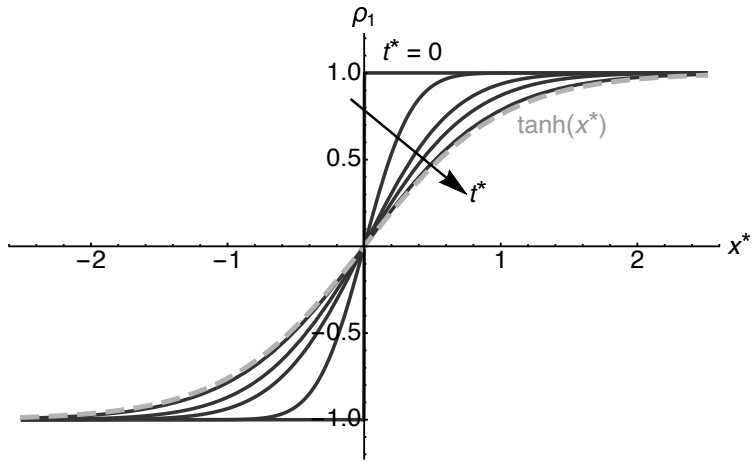}
    \caption{Shock formation process as given by Eq. \eqref{eq:solutionRho1}. Flow moves from left to right. The times depicted are $t^* = 0, 0.1, 0.5, 1$ and $3$. The shock is essentially fully developed for $t^*>3$.} 
    \label{fig:TempEvolution}
\end{figure}

We study the shock formation problem arising from a Heaviside-step initial condition 

\begin{equation}
    \rho_{1}=\left\{ \begin{array}{cc}
    -1 & \text{for }x^{*}<0,\\
    1 & \text{for }x^{*}>0.
\end{array}\right.
    \label{eq:initialCondition}
\end{equation}
This choice for initial condition is relatively unusual because $\rho _1$ refers to the first-order density perturbation, which we set to be nonzero upstream.
This choice, however, is motivated by the fact that the  solution to Eq. \eqref{eq:Burgers Cond} satisfying Eq. \eqref{eq:initialCondition} takes the compact form\cite{hu1972collisional}

\begin{equation}
    \rho _1\left(t^*,x^*\right) = \dfrac{\left[ 1+ \text{erf}\left(\dfrac{t^*+x^*}{\sqrt{2t^*}}\right)\right]\exp\left(2x^*\right)- 1- \text{erf}\left(\dfrac{t^*-x^*}{\sqrt{2t^*}}\right)}
    {\left[ 1+ \text{erf}\left(\dfrac{t^*+x^*}{\sqrt{2t^*}}\right)\right]\exp\left(2x^*\right)+ 1+ \text{erf}\left(\dfrac{t^*-x^*}{\sqrt{2t^*}}\right)},
    \label{eq:solutionRho1}
\end{equation}
shown in Fig. \ref{fig:TempEvolution}.
The error function is defined as usual,

\begin{equation}
    \text{erf}\left(t\right)=\dfrac{2}{\sqrt{\pi}}\int _0  ^t \exp \left(-\tau^2 \right)\text{d}\tau.
    \label{eq:error}
\end{equation}
It can be seen that the density profile approaches a steady shock structure asymptotically in time, given by
\begin{equation}
    \left. \rho _1\right|_{t^*\rightarrow \infty} = \tanh\left(x^*\right).
    \label{eq:Tanh}
\end{equation}

Returning to dimensional variables, the following density profile is therefore established to the first order in $\epsilon$:

\begin{equation}
    \rho\left(t,x\right)_{\text{cond.}}=\rho _0 \left\lbrace 1+\epsilon \tanh \left[\dfrac{\gamma\left(\gamma +1\right)}{2\left(\gamma -1\right)^2} \dfrac{\epsilon}{l_c}\left(x+c_0 t\right)\right] \right\rbrace.
    \label{eq:SolutionDensDim}
\end{equation}
This is the shock structure generated by a pressure imbalance $\Delta p/p_0=2\gamma\epsilon$ or the impulsive motion of a piston with velocity $u/c_0=2\epsilon$.
In the absence of radiative effects, the characteristic shock width is $\mathcal{L}_{\text{cond.}}=l_c/\epsilon$, and the shock structure is developed in a characteristic time $\mathcal{T}_{\text{cond.}}= l_c/c_0\epsilon ^2$.

\subsection{\label{subsec:Radiation}Weak shocks with radiative effects}

We now consider how radiation alters the structure of the shock. 
It is first necessary to precondition the system according to the scaling of the Boltzmann number and optical thickness with respect to the shock strength $\epsilon$. 
For this purpose, we propose to rewrite it in the following form 

\begin{equation}
    \left(\dfrac{\tau}{\epsilon}\right)^2 \sim O\left(\epsilon ^ a\right),\quad \dfrac{\epsilon\text{Bo}}{\tau}\sim O\left(\epsilon ^b\right),
    \label{eq:BoTauab}
\end{equation}
with $a,b$ integers. 
The first scaling characterizes the behavior of the radiative transfer equation \eqref{eq:radiationNorm}, while the second measures the effect of radiation on the energy equation \eqref{eq:energyNorm}.

Positive values of $a$ correspond to optically thin regimes, as the radiative transfer equation yields 

\begin{equation}
    \dfrac{\partial^2 T_{R_j}}{\partial x^{\prime 2}}=0 \quad  \text{  for } j\leq a;
    \label{eq:a OptThin}
\end{equation}
that is, radiation temperature remains constant up to the order $\epsilon ^a$. 
Negative values of $a$ correspond to optically thick regimes, resulting in

\begin{equation}
     T_j = T_{R_j} \quad  \text{  for } j\leq |a|;
    \label{eq:a OptThick}
\end{equation}
which implies that matter and radiation temperatures are in equilibrium up to the order $\epsilon ^{|a|}$. 
The radiative transfer equation also relates in this case temperatures imbalance of higher orders, $\left.\left( T_{j} - T_{R_{j}} \right)\right|_{j>|a|}$, to derivatives of the radiation temperature at orders $\epsilon^{j-|a|}$ and lower. 

The integer $b$ assesses the order at which radiation affects the hydrodynamics through the energy equation \eqref{eq:energyNorm}. 
More precisely, the term $\left( T_{b+1} - T_{R_{b+1}} \right)$ enters the energy equation at order $\epsilon$, while $\left( T_{b+2} - T_{R_{b+2}} \right)$ comes in play at order $\epsilon ^2$.
In addition, positive values of $b$ imply equilibrium of matter and radiation temperatures up to order $b$, since no other term in the energy equation could support such imbalance. 

\begin{figure}[H]
    \centering
    \includegraphics[width=0.7\linewidth]{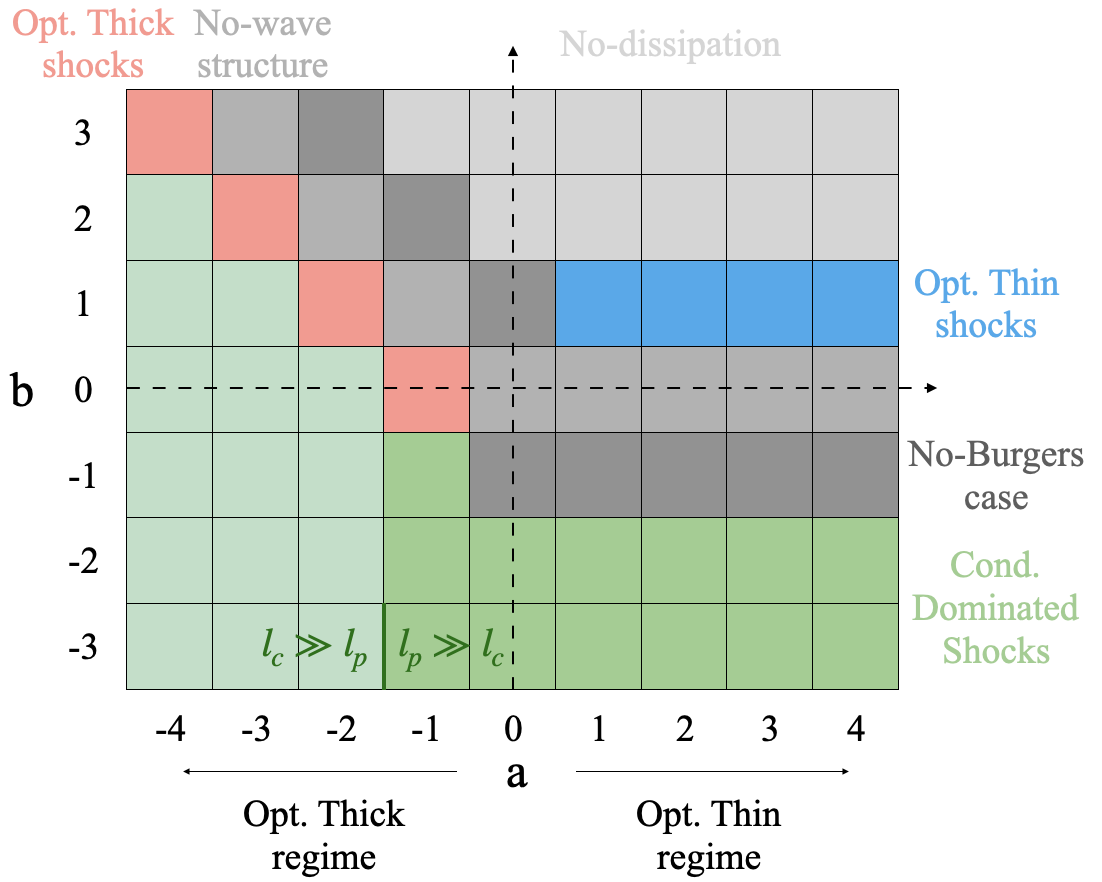}
    \caption{Phenomenology of shock structure depending on the scaling $ \left(\tau/\epsilon\right)^2 \sim O\left(\epsilon ^ a\right)$ and $\epsilon\text{Bo}/\tau\sim O\left(\epsilon ^b\right)$.} 
    \label{fig:Chart}
\end{figure}

The phenomenology of solutions is summarized in Fig. \ref{fig:Chart}.
It follows from the analysis of $b$ that radiation does not affect the shock structure for $b\leq-2$, since the procedure is closed at order $\epsilon ^2$ of the governing equations without radiative effects entering the formulation. 
The radiative flux is too low in this case (the Boltzmann number too high) to affect the shock profile, which is mediated by thermal conduction and given by Eq. \eqref{eq:SolutionDensDim}. 
These conduction-dominated solutions are shown in green in Fig. \ref{fig:Chart}. 
The optical thickness then becomes $\tau\sim l_c/l_p$, which sets the value of $a$.
Particularly, the physical condition $l_c\ll l_p$ restricts the validity of conduction-dominated shock solutions to $a\geq -1$.

\subsubsection{\label{subsubsec:Thick}Optically thick solutions}

We next evaluate the possibility of radiation-dominated, optically thick solutions $(a<0)$. 
The radiative transfer equation establishes that the term $\left.\partial ^2 T_{R_{j}}/\partial x^{\prime 2}\right|_{j=b+1-|a|}$ appear in the order $\epsilon$ of the energy equation. 
Therefore, if $b = |a|$, such order does not conform a homogeneous system in the first spatial derivatives of the perturbed variables, and the  eigenvalue $v_0$ cannot be retrieved.
Physically, this entails that perturbations are not of a wave-traveling type, and should not be sought under the stretching proposed in Eq. \eqref{eq:stretching}.
This case is depicted in gray in Fig. \ref{fig:Chart}.
Following a similar logic, the term $\left.\partial ^2 T_{R_{j}}/\partial x^{\prime 2}\right|_{j=b+2-|a|}$ appears in the order $\epsilon ^2$ of the energy equation. 
It follows then that, if $b\leq |a|-2$, radiation does not contribute to this order, and we fall again in conduction-dominated solutions.

\begin{figure}[H]
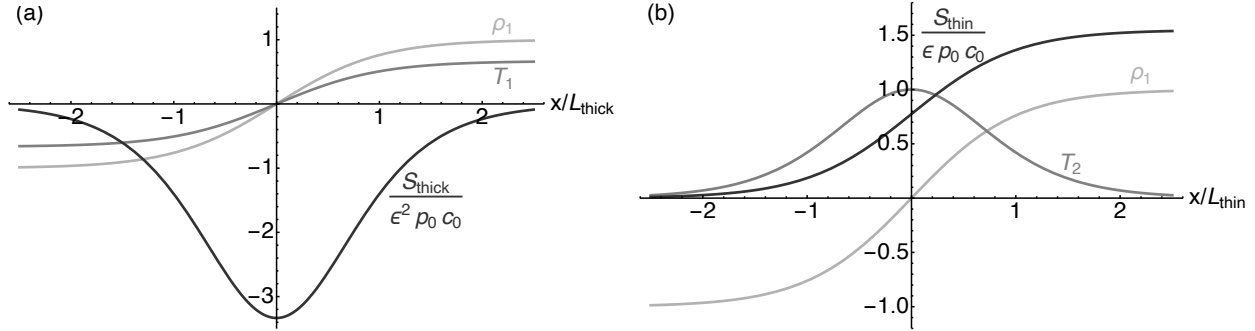

    \centering
    \begin{subfigure}[b]{0.49\textwidth}
        \includegraphics[width=\textwidth]{Images/StructureThick.png}
    \end{subfigure}
    \hfill
    \begin{subfigure}[b]{0.49\textwidth}
        \includegraphics[width=\textwidth]{Images/StructureThin.png}
    \end{subfigure}
    \caption{Analytical profiles for (a) optically thick and (b) optically thin radiation-dominated weak shocks. Flow is moving from left to right. The value $\gamma = 5/3$ has been assumed. The first non-vanishing term in temperature perturbation is shown for each case.}
    \label{fig:StructureShocks}
\end{figure}

A plausible option for optically thick shocks is $b=|a|-1$, painted in orange. 
In this case, perturbations travel at the isentropic speed of sound, since the order $\epsilon$ yields the same system as Eq. \eqref{eq:Order1}.
First-order perturbations satisfy the same relation conditions, that is, $u_1=-v_0\rho_1$, $T_1=T_{R_1}=\left(\gamma -1\right)\rho_1$.
Radiation manifests as a heat-conduction dissipative process, modifying the Burgers-type equation derived from order $\epsilon ^2$, which becomes

\begin{equation}
    \dfrac{\partial \rho_1}{\partial t^{\prime}}-\dfrac{\gamma +1}{2}\rho_1\dfrac{\partial \rho_1}{\partial x^{\prime}}=\dfrac{\left(\gamma -1\right)^2}{2\gamma}\left(\kappa + \dfrac{4}{3\tau \text{Bo}}\right)\dfrac{\partial ^2 \rho _1}{\partial x^{\prime 2}}.
    \label{eq:Burgers Thick}
\end{equation}
We notice that $b=|a|-1$ implies $\tau\sim 1/\text{Bo}$, hence the new term introduced in Eq. \eqref{eq:Burgers Thick} is of the order of unity. 
Comparing then the two dissipative processes, we obtain 

\begin{equation}
    \dfrac{\kappa}{4/\left(3\tau\text{Bo}\right)}\sim \text{Bo}\dfrac{l_c}{l_p},
    \label{eq:Comparison dissipation}
\end{equation}
which we assume to be small since positive $b$ values imply a somewhat moderate Bo at most.
These shocks are therefore radiation dominated and the following density profile is established in a time scale $\mathcal{T}_{\text{thick}}= l_p/\text{Bo}c_0\epsilon^2$,

\begin{equation}
    \rho\left(t,x\right)_{\text{thick}}=\rho _0 \left\lbrace 1+\epsilon \tanh \left[\dfrac{3\gamma\left(\gamma +1\right)}{8\left(\gamma -1\right)^2} \dfrac{\epsilon\text{Bo}}{l_p}\left(x+c_0 t\right)\right] \right\rbrace.
    \label{eq:SolutionDensDimThick}
\end{equation}
The scaled length of the shock becomes $L\sim l_p/\text{Bo}$, which verifies the initial estimate $\tau\sim 1/\text{Bo}$.
The radiation energy flux $S$ can be expressed in this case as

\begin{equation}
    \dfrac{S_{\text{thick}}}{p_0c_0}=-\epsilon^2\dfrac{\gamma\left(\gamma+1\right)}{2\left(\gamma -1\right)}\left(1-\rho_1^2\right),
    \label{eq:SThick}
\end{equation}
which is always negative and vanishes both far upstream and far downstream. 
It has been depicted in Fig. \ref{fig:StructureShocks}(a). 
The radiation emitted from the compressed material is therefore used to heat up the colder incoming flow.

We emphasize that optically thick, radiation-dominated shocks can only take place if $b\geq 0$, which is equivalent to $\text{Bo}\lesssim \epsilon^{-1/2}$. 
A lower Boltzmann number results in thicker shocks. 
Particularly, for $\text{Bo}\sim \epsilon^{r/2}$, with $r\geq-1$, the shock width scales as $\mathcal{L}_{\text{thick}}= l_p/\epsilon\text{Bo}\sim l_p/\epsilon^{1+r/2}$, and matter and radiation temperatures are in equilibrium up to the order $\epsilon ^{r+2}$.
Contrary to this, a shock of strength $\epsilon$ emitted in a medium with $\text{Bo}\gg  \epsilon^{-1/2}$ will not be mediated by radiation. 

\subsubsection{\label{subsubsec:Thick}Optically thin solutions}

We now consider optically thin regimes described by $a>0$, in which radiation temperature remains constant up to the order $\epsilon ^a$.  
This implies that $T_1$ must be equal to zero if $b>0$, since the energy equation enforces temperature equilibrium up to the order $\epsilon ^b$. 
The first order in $\epsilon$ of mass and momentum conservation equations  then becomes uncoupled from the energy equation, resulting in perturbations travelling at the isothermal speed of sound, $v_0 = 1/\sqrt{\gamma}$.
However, not all values of $b$ are valid. 
Following a similar reasoning, the second term of the temperature expansion $T_2$ is also zero when $b > 1$.
The resulting evolutionary equation for $\rho_1$ can then be obtained from mass and momentum conservation alone and lacks a dissipation mechanism: $\partial \rho_1 /\partial t=\rho_1\partial \rho_1 / \partial x$. 
This is an indication that a too thin optical thickness is estimated \emph{a priori} when taking $b>1$. 
This no-dissipation case is extensible to non-positive $a$ values as long as $a > 1 - b$, and is marked in light gray in Fig. \ref{fig:Chart}.

The only option for perturbations to propagate at the isothermal speed of sound is therefore $b=1$, shown in blue in Fig. \ref{fig:Chart}.
The evolutionary equation for density can still be obtained from mass and momentum conservation alone, which at the order $\epsilon ^2$ read

\begin{equation}
    v_0\dfrac{\partial \rho_2}{\partial x{^\prime}} + \dfrac{\partial u_2}{\partial x{^\prime}}
    =
    -\dfrac{\partial \rho_1}{\partial t{^\prime}} +2v_0\rho_1\dfrac{\partial \rho_1}{\partial x{^\prime}},
    \label{eq:Mass2Thin}
\end{equation}
and
\begin{equation}
    \dfrac{1}{\gamma}\dfrac{\partial \rho_2}{\partial x{^\prime}} + v_0\dfrac{\partial u_2}{\partial x{^\prime}}
    =
    v_0\dfrac{\partial \rho_1}{\partial t{^\prime}} -\dfrac{1}{\gamma}\dfrac{\partial T_2}{\partial x{^\prime}},
    \label{eq:Mom2Thin}
\end{equation}
respectively.
However, now there is a term involving $T_2$, which can be related to $\rho _1$ using the energy equation at the previous order $\epsilon$.
The latter balances the compressional pdV work with radiative cooling, reading 

\begin{equation}
    \dfrac{\partial u_1}{\partial x{^\prime}}=-\dfrac{4\tau}{\text{Bo}}T_2 \Longrightarrow 
    T_2=\dfrac{\text{Bo}}{4\tau}v_0\dfrac{\partial \rho_1}{\partial x{^\prime}}.
    \label{eq:T2}
\end{equation}
Bearing this into account, and taking $v_0=1/\sqrt{\gamma}$ in Eqs. \eqref{eq:Mass2Thin}, and \eqref{eq:Mom2Thin}, a Burgers-type equation can be derived for the perturbed density,

\begin{equation}
    \dfrac{\partial \rho_1}{\partial t^{\prime}}-\dfrac{1}{\sqrt{\gamma}}\rho_1\dfrac{\partial \rho_1}{\partial x^{\prime}}
    =
    \dfrac{\text{Bo}}{8\gamma\tau}\dfrac{\partial^2\rho_1}{\partial x^{\prime 2}}.
    \label{eq:Burgers Thin}
\end{equation}
This equation governs the density profile of optically thin, radiation-dominated shocks. 
The assumption $b=1$ implies $\tau \sim \text{Bo}$, hence the right-hand-side term in Eq. \eqref{eq:Burgers Thin} is of the order of unity. 
The fact that the only dissipative mechanism present is of a radiative nature is a consequence of the shock evolving isothermally to the first order. 
The density profile tends asymptotically in time to
\begin{equation}
    \rho\left(t,x\right)_{\text{thin}}=\rho _0 \left\lbrace 1+\epsilon \tanh \left[4\sqrt{\gamma}\dfrac{\epsilon}{l_p\text{Bo}}\left(x+\dfrac{c_0}{\sqrt{\gamma}} t\right)\right] \right\rbrace,
    \label{eq:SolutionDensDimThin}
\end{equation}
which is established in a characteristic time scale $\mathcal{T}_{\text{thin}}= l_p\text{Bo}/c_0\epsilon^2$.
The scaled length $L\sim l_p\text{Bo}$ verifies the initial estimate $\tau \sim \text{Bo}$. 

We have shown that optically thin, radiation-dominated shocks can only take place if the Boltzmann number of the medium in which they propagate satisfies $\text{Bo}\lesssim \epsilon^{3/2}$ (equivalent to $a\geq 1$ with $b=1$). 
We notice that a small Boltzmann was also necessary for the formation of an optically-thin relaxation layer in strong stationary shocks, see Fig. \ref{fig:OTRsketch}(b). 
Interestingly, in this regime, a lower Bo results in thinner shocks, since the shock width scales as $\mathcal{L}_{\text{thin}}= l_p\text{Bo}/\epsilon$.
In these shocks, matter is compressed quasi-isothermally and higher densities can be achieved for a fixed compression level.
Effectively, one can retrieve $\Delta \rho /\rho_0 = \Delta p / p_0$ to the first order in this regime as opposed to $\Delta \rho /\rho_0 = \left(1/\gamma\right)\Delta p/p_0$ in  optically thick weak shocks, where matter is compressed quasi-adiabatically.

The radiation energy flux $S$ plays a key role in the structure of these shocks. 
It can easily be proven that optically thin shocks launched in a medium where $\text{Bo}\sim \epsilon^{r/2}$ have $T_{R_{j}}=\text{const.}$ for $j\leq r-1$.  
Bearing this into account and after some algebra, the following expression for $S$ is obtained,

\begin{equation}
    \dfrac{S_{\text{thin}}}{p_0c_0}=\dfrac{\epsilon}{\sqrt{\gamma}}\left(1+\rho_1\right).
    \label{eq:SThin}
\end{equation}
The radiation energy flux is therefore positive and does not vanish far downstream in the compressed material, see Fig. \ref{fig:StructureShocks}(b). 
The incoming flow essentially radiates away any internal energy gained during its compression.  
The radiation energy flux advects this energy away from the shock, preventing the heating of the matter. 

It should be noted that optically thin shocks cannot be sustained indefinitely in time, as the outflowing radiation flux would eventually be absorbed in a larger scale compared to the shock thickness. 
These shock solutions should therefore be understood as transitory, existing in a time scale $t\ll\mathcal{T}_{\text{thick}}$ before a wider, optically thick shock structure embeds them. 

There are some remaining cases that we have not discussed yet, which concern $\lbrace b = 0$, $a \geq 0\rbrace$, $\lbrace b = -1$, $a\geq 0\rbrace$, and $\lbrace b = |a|+1$, $a \leq 0\rbrace$.
The former presents a term in $T_1$ in the order $\epsilon$ of the governing equations, hence we retrieve the no-wave case since the eigenvalue $v_0$ cannot be determined.
The latter two correspond to perturbations traveling at the isentropic and isothermal speed of sound, respectively, but the governing equation for the perturbed density does not possess a Burgers-type structure.
These cases, shown in dark gray in Fig. \ref{fig:Chart}, might support solutions of a traveling-type nature different to shock waves, being as such out of the scope of this paper and left for future analysis. 

\subsection{\label{subsec:Picture}Comprehensive picture of shock formation}

We have identified three different solutions for a shock profile: conduction-dominated, optically thin radiation-dominated, and optically thick radiation-dominated. 
Their optical thicknesses are $\tau_{\text{cond.}} = l_c/l_p$, $\tau_{\text{thin}} = \text{Bo}$ and $\tau_{\text{thick}} = 1/\text{Bo}$, respectively.
Optically thick solutions can develop for $\text{Bo}\lesssim \epsilon ^{-1/2}$, whereas optically thin solutions require $\text{Bo}\lesssim \epsilon ^{3/2}$.
Therefore, if the latter inequality is satisfied, the three shock structures are possible, but present different formation time $\mathcal{T}$. 
We can establish in this case $\mathcal{T}_{\text{cond.}}\ll \mathcal{T}_{\text{thin}}\ll \mathcal{T}_{\text{thick}}$.

\begin{figure}[H]
    \centering
    \includegraphics[width=0.8\linewidth]{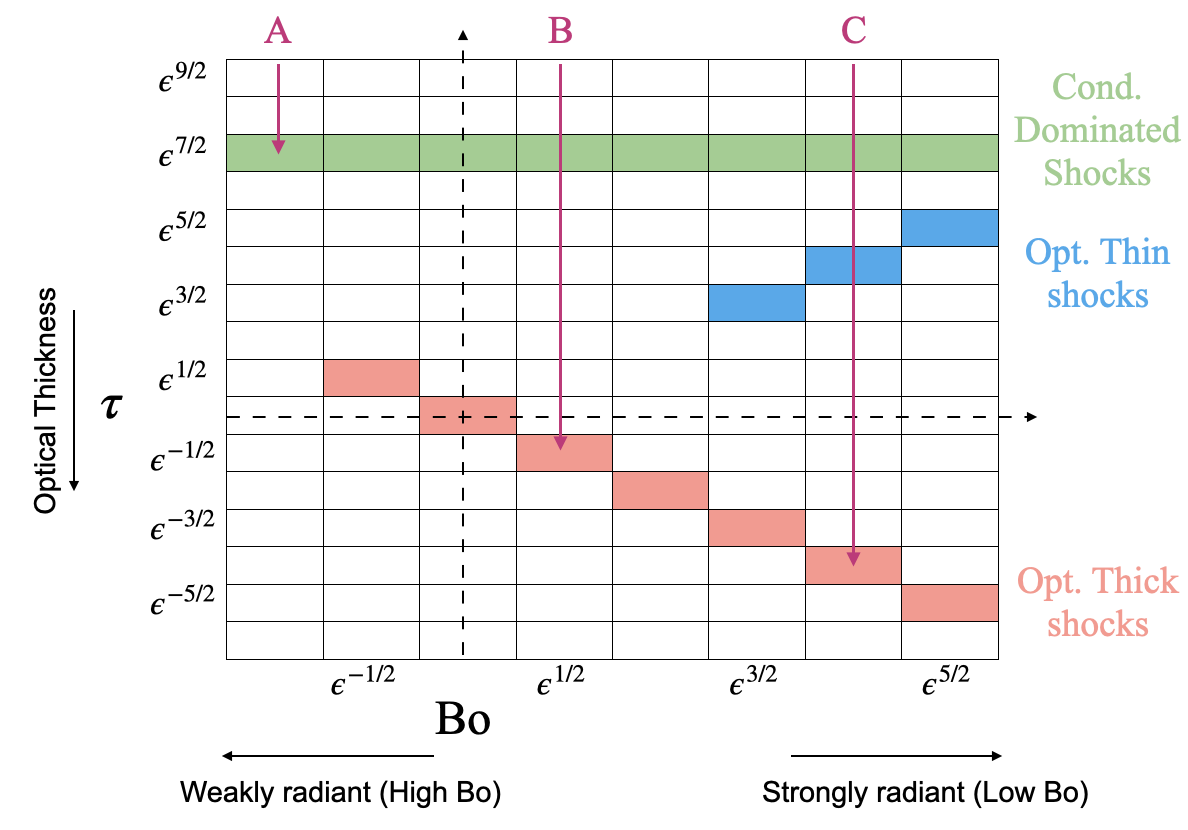}
    \caption{Phenomenology of the formation of weak shocks of strength $\epsilon\ll 1$ in radiative media as a function of its Boltzmann number. The arrows in magenta represent the temporal growth of the shock thickness until the main dissipation mechanism is established. In this chart, we have chosen $l_c/l_p\sim \epsilon ^{7/2}$, which sets the optical thickness of conduction-dominated shocks.} 
    \label{fig:Chart2}
\end{figure}

We now analyze the formation of a shock structure after two radiative plasmas with different pressures are put in contact. 
Three different scenarios are possible depending on the scaling of the Boltzmann number with the shock strength $\epsilon$, as shown in Fig. \ref{fig:Chart2}.
As the shock starts to form, its optical thickness increases. 
First, conduction becomes predominant, and a density profile given by Eq. \eqref{eq:SolutionDensDim} is established in a time scale $\mathcal{T}_{\text{cond.}}$. 
The optical thickness of this profile is still very small, and matter and radiation are not in equilibrium across the shock. 
However, if $\text{Bo}\gg \epsilon ^{-1/2}$, the compressed material does not radiate enough to modify the structure of the shock, which stays conduction-dominated.
This is the case labeled $A$.
For lower Boltzmann numbers, radiative effects come into play and alter the shock structure. 
Matter and radiation tend to equilibrate in a longer time scale $\mathcal{T}_{\text{thick}}$, resulting in the formation of a thicker, radiation-dominated shock profile as given by Eq. \eqref{eq:SolutionDensDimThick}.
This is case $B$ in the figure. 
For even lower Boltzmann numbers satisfying $\text{Bo}\ll \epsilon ^{3/2}$, an intermediate regime arises in a time scale $\mathcal{T}_{\text{thin}}$ where matter and radiation are still not in equilibrium, but the shock profile is mediated by radiation, Eq. $\eqref{eq:SolutionDensDimThin}$.
As discussed earlier, this solution cannot be maintained indefinitely, and always degenerates into the optically thick structure asymptotically in time. This is illustrated by case $C$.

Of similar interest is to evaluate the nature of the structure developed when the characteristic of the medium is fixed and one varies the strength of the shock. 
According to the phenomenology in Fig. \ref{fig:Chart2}, in a medium where $\text{Bo}\gg 1$, strong enough shocks satisfying $\epsilon \gg \text{Bo}^{-2}$ can only be mediated by thermal conduction. 
Optically thick, radiation-dominated shocks develop for weaker shock intensities.
The situation changes in a strongly radiant medium where $\text{Bo} \lesssim 1$, since all the weak shocks become optically thick asymptotically in time. 
In this case, optically thin, radiation-dominated transitory solutions develop only for a strong enough shock satisfying $\epsilon \gtrsim \text{Bo}^{2/3}$. 
This is the limit of interest to experience overcompression in Z-pinch liners, and possible applications are discussed in Sec. \ref{Sec: conclusions}.

\section{\label{Sec:numerical}Numerical simulations}

The theoretical analysis has been supported with numerical simulations of shock formation using the radiation-hydrodynamics code \emph{FLASH}.\cite{fryxell2000flash, Tzeferacos2015} 
It is a publicly-available, multi-physics, adaptive mesh refinement, finite-volume Eulerian hydrodynamics and MHD code, developed at the University of Rochester by the Flash Center for Computational Science (for more information on the \emph{FLASH} code, visit: \protect\url{https://flash.rochester.edu}).
The Flash Center has added extensive high-energy-density physics and extended-MHD capabilities \cite{Tzeferacos2015} that make \emph{FLASH} an ideal tool for bench-marking our results. 
\emph{FLASH} has been validated through extensive benchmarks and code-to-code comparisons,~\cite{Fatenejad2013, Orban2013, Orban2022, Sauppe2023} as well as through direct application to numerous plasma-physics experiments.~\cite{Falk2014, Yurchak2014, Meinecke2014, Li2016, Tzeferacos2018, Rigby2018, Gao2019, Bott2021, Meinecke2022} 
For pulsed-power experiments, \emph{FLASH} has been able to reproduce past analytical models,\cite{Slutz2001} has been applied in the modeling of capillary discharge plasmas\cite{Cook2020} and staged Z-pinch implosions, \cite{Hansen2023PoP} and is being validated against gas-puff experiments at CESZAR.\cite{Conti2020PRAB}

\begin{table}[h]
    \centering
    \setlength{\tabcolsep}{5pt} 
    \begin{adjustbox}{width=\textwidth}
        \begin{tabular}{ c || c | c | c | c | c | c | c | c | c }
            Fig. & $\rho _0$& $T_0$& $\epsilon$ & Bo & Ma & Conductivity & Opacity& $L_c$& $t_c$\\ 
             & [mg/cm$^3$]& [eV] & & & & [cm$^{-1}$s$^{-1}$] & [cm$^{-1}$] &[cm] & [s]\\ 
            \hline
            \hline
            6(a)       & 12.6  & 10 & 0.03   & 33.3 [$\epsilon^{-1}$]  & 1.04 & 6.03$\times 10^{22}$ & 678                 & $f_{1\gamma}\mathcal{L}_{\text{cond.}} =  4.74\times 10^{-6}$ & $f_{2\gamma}\mathcal{T}_{\text{cond.}} =  2.10\times 10^{-11}$\\ 
            6(b)       & 20.6  & 100   & 0.03      & 0.17 [$\epsilon^{1/2}$]  & 1.04 & 1.90$\times 10^{25}$ & 0.579               & $g_{1\gamma}\mathcal{L}_{\text{thick}} =  88.7$ & $g_{2\gamma}\mathcal{T}_{\text{thick}} =  1.24\times 10^{-4}$  \\
            6(c),(d)  & 0.107 & 100 & 0.03 & $9\times10^{-4}$ [$\epsilon^{2}$] & 0.80 & 1.90$\times 10^{25}$ & 1.56$\times 10^{-5}$ & $h_{1\gamma}\mathcal{L}_{\text{thin}} =  372$ & $h_{2\gamma}\mathcal{T}_{\text{thin}} =  8.95\times 10^{-4}$\\ 
            \hline
            7(a)            & 1.19  & 100  & 0.05   & 0.01 [$\epsilon^{3/2}$] & 1.07 & 1.90$\times 10^{25}$ & 1.93$\times 10^{-3}$ & $h_{1\gamma}\mathcal{L}_{\text{thin}} =  20.1$ & $h_{2\gamma}\mathcal{T}_{\text{thin}} =  290\times 10^{-5}$\\
            7(b)        & 1.19  & 100    & 0.05   & 0.01 [$\epsilon^{3/2}$]   & 1.07 & 1.90$\times 10^{25}$ & 1.93$\times 10^{-3}$ & $g_{1\gamma}\mathcal{L}_{\text{thick}} =  2.76\times 10^{5}$ & $g_{2\gamma}\mathcal{T}_{\text{thick}} =  0.232$\\ 
            \hline
            8(a)         & 12.6  & 10   & 0.03     & 33.3 [$\epsilon^{-1}$]  & 1.04 & 6.03$\times 10^{22}$ & 678                 & $g_{1\gamma}\mathcal{L}_{\text{thick}} =  3.93\times 10^{-4}$ & $g_{2\gamma}\mathcal{T}_{\text{thick}} =  1.74\times 10^{-9}$\\
            8(b)          & 20.6  & 100   & 0.03   & 0.17 [$\epsilon^{1/2}$]  & 1.04 & 1.90$\times 10^{25}$ & 0.579               & $h_{1\gamma}\mathcal{L}_{\text{thin}} =  1.93$ & $h_{2\gamma}\mathcal{T}_{\text{thin}} =  4.65\times 10^{-6}$  \\
        \end{tabular}
    \end{adjustbox}
    \caption{Parameters of the simulations performed. The upstream conditions $\rho _0$, $T_0$ and shock strength $\epsilon$ are the input quantities, with the rest of the variables deriving from them. The Boltzmann number Bo, conductivity and opacity are in accordance with the numerical formulas \eqref{eq:BoltzmannNum} -- \eqref{eq:lpNum}. The relative scaling of Bo with respect to $\epsilon$ is shown in brackets. Ma refers to the isentropic Mach number of the incoming flow. $L_c$ and $t_c$ refer to the characteristic length and formation time in each simulation case. They are determined by the main dissipative mechanism of the regime of interest. The preceding multiplying factors ensure that the corresponding governing equation in dimensionless form becomes the Burgers equation \eqref{eq:Burgers Cond Norm} regardless of the regime. They read $f_{1\gamma} = 2(\gamma-1)^2/\gamma(\gamma+1)$, $f_{2\gamma} = 2f_{1\gamma}/(\gamma +1)$, $g_{1\gamma} = 4f_{1\gamma}/3$, $g_{2\gamma} = 4f_{2\gamma}/3$, $h_{1\gamma} = 1/(4\sqrt{\gamma})$, $h_{2\gamma} = 1/4$.}
    \label{tab:CaseDescription}
\end{table}

\begin{figure}[H]
    \centering
    \begin{subfigure}[b]{0.49\textwidth}
        \includegraphics[width=\textwidth]{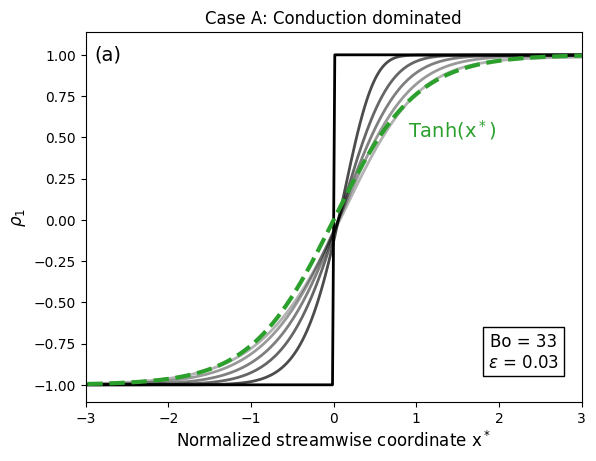}
        \includegraphics[width=\textwidth]{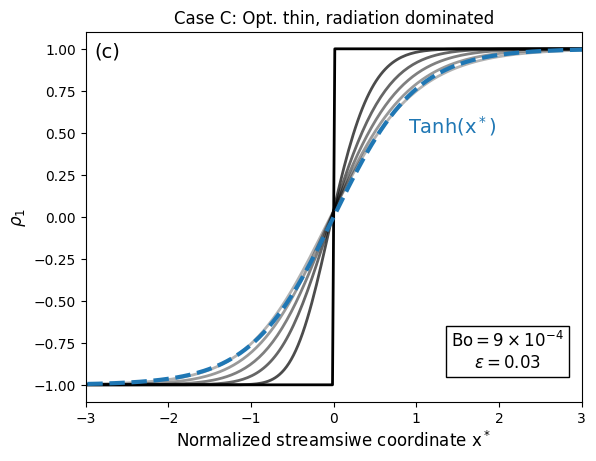}
    \end{subfigure}
    \hfill
    \begin{subfigure}[b]{0.49\textwidth}
        \includegraphics[width=\textwidth]{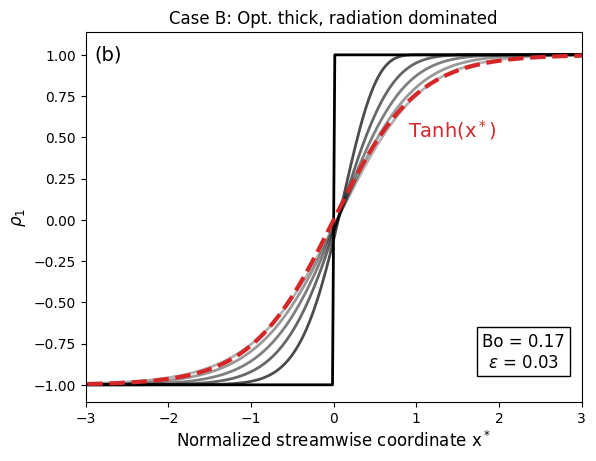}
        \includegraphics[width=\textwidth]{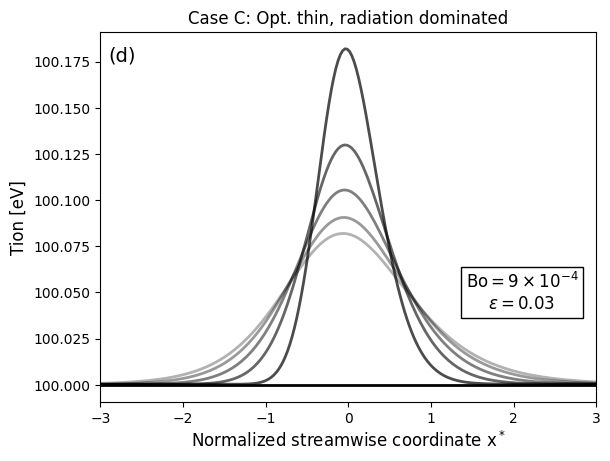}
    \end{subfigure}
    \caption{Evolution of the normalized perturbed density (a)-(c), and matter temperature (d) from three different \emph{FLASH} simulations. They correspond to cases A, B, and C sketched in Fig. \ref{fig:Chart2}, characterized by Bo = $\epsilon ^{-1}, \epsilon ^{1/2}, \epsilon ^2$, respectively, and particularized for $\epsilon = 0.03$. The characteristic lengths and times provided in Table \ref{tab:CaseDescription} have been used to normalize the streamwise coordinate and evolution time. In every panel, darker to lighter grays correspond to normalized times $t/t_c = 0, 0.2, 0.5, 1, 2$ and 5. Density profiles are compared to the fully-formed analytical solution in dashed lines.} 
    \label{fig:NumericalShocks}
\end{figure}

Our simulation setup consists of two plasmas with different pressures put in contact at $x = 0$.
We consider inviscid, fully ionized hydrogen plasmas satisfying a gamma-law equation of estate with an adiabatic ratio $\gamma = 5/3$.
Thermal conductivity given by Braginskii, and we have taken a Coulomb logarithm value equal to 10.
Consistently with the theoretical model, diffusion approximation for the radiant flux is assumed, and we have used gray opacities given by the inverse of the photon mean-free-path expression in Eq. \eqref{eq:lpNum} for both Rosseland and Planck.
Matter and radiation are initially at equilibrium at each side of the discontinuity. The density jump is related to the shock strength $\epsilon$ through $\rho_1 /\rho_0 =1+ 2\epsilon$, with the subscripts 1, 0 denoting the compressed and uncompressed regions, respectively.
The fluid velocity in the latter has been chosen so that the shock remains at rest in the simulation domain in each case. 
Unless otherwise specified, the jump in the rest of the variables is dictated by the Rankine-Hugoniot conditions. 
Table \ref{tab:CaseDescription} details the upstream conditions and shock strength for each of the simulation cases plotted below, as well as the corresponding dimensionless numbers, characteristic lengths $L_c$ and formation times $t_c$. 
The computational domain in every case spans 20$\times L_c$, and we have used 720 grid points. 
The particular upstream conditions chosen ensure a large thermal-to-radiation pressure ratio $\alpha$, required for the analysis. 
The most limiting case would be Fig. \ref{fig:NumericalShocks}(c), where $\alpha \approx 5$.

The formation of shocks of strength $\epsilon = 0.03$ in plasmas with different Boltzmann numbers is shown in Fig. \ref{fig:NumericalShocks}. 
We have chosen cases A, B, and C conceptualized in Fig. \ref{fig:Chart2}, where Bo scales as $\epsilon^{-1}$, $\epsilon^{1/2}$, and $\epsilon^{2}$, respectively. 
Each presents different characteristic lengths and formation times, determined by the mechanisms of thermal conduction, radiation in optically thick regime and radiation in optically thin regime, respectively.
Panels \ref{fig:NumericalShocks}(a)-(c) show the normalized perturbed density.
It can be seen that by $t =5\times t_c $, the shock profile has relaxed to a hyperbolic tangent function, demonstrating excellent agreement with the analytical expressions provided for each regime, Eqs. \eqref{eq:SolutionDensDim}, \eqref{eq:SolutionDensDimThick} and \eqref{eq:SolutionDensDimThin}.
We remark that we deemed necessary to artificially increase the ion-electron coupling frequency in Case A due to the fast formation time of conduction-dominated shocks. 
Although not shown in the figures, we notice that radiation temperature quickly flattens out in Case A, while it remains in equilibrium with the matter temperature at all times in Case B.
The density and temperature profiles developed in this opt. thick case match the analytical profiles sketched in Fig. \ref{fig:StructureShocks}(a).
The evolution of matter (ion) temperature in the opt. thin Case C is shown in Fig. \ref{fig:NumericalShocks}(d).
Unlike in the other regimes, the theoretical analysis has proven that the shock forms in quasi-isothermal conditions, requiring the simulation to begin with a homogeneous temperature profile. 
The Mach number of the incoming flow is derived in this case from the momentum equation alone, reading Ma $= \left[ (1 + 2\epsilon)/\gamma \right] ^{1/2}$. 
The temperature profile peaks at the center of the shock, broadening and diminishing its maximum as the shock evolves. 
It experiences a second-order increase, $\Delta T/T \sim O(\epsilon^2)$, as predicted by the theoretical analysis and in agreement with the results for $T_2$ provided by Eq. \eqref{eq:T2} and sketched in Fig. \ref{fig:StructureShocks}(b).

\begin{figure}[H]
    \centering
    \includegraphics[width=0.49\textwidth]{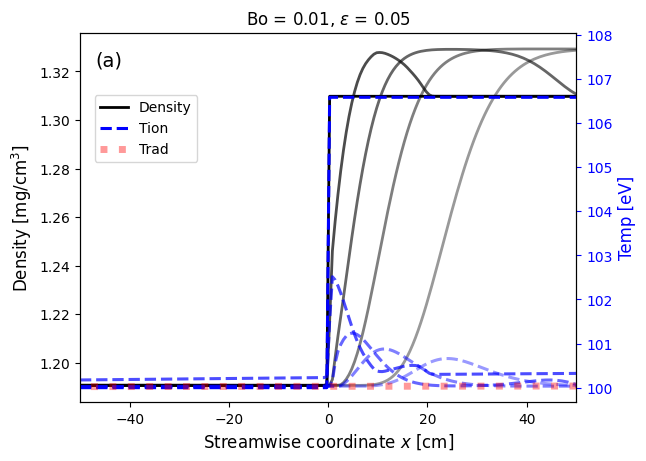}
    \hfill
    \includegraphics[width=0.49\textwidth]{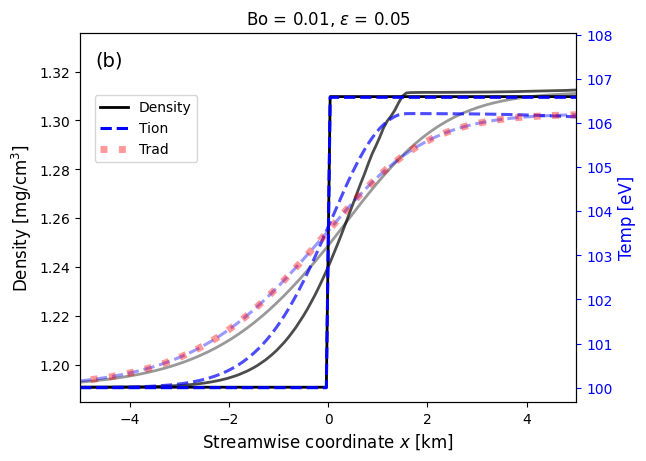}
    \caption{Evolution of density and temperature from two different  \textit{FLASH} simulations of the same case $\epsilon = 0.05$ and Bo = 0.01 ($\sim \epsilon ^{3/2}$). Each simulation is characterized by a different computational length and time, corresponding to (a) optically thin, and (b) optically thick regimes. Notice that the scale of the streamwise coordinate changes in each panel. Darker to lighter curves depict increasing simulation time: (a) $t/t_c = 0, 0.02, 0.05, 0.1$ and 0.2. (b)  $t/t_c = 0, 0.2$ and 2. The normalizing time $t_c$ is different in each panel and provided in Table \ref{tab:CaseDescription}. Radiation temperature at the latest plotting time is shown in dashed red lines.}
    \label{fig:OptThinAnalysis}
\end{figure}

Figure \ref{fig:OptThinAnalysis} depicts a shock developing in conditions satisfying Bo $\sim O(\epsilon ^{3/2})$, where the optically thin regime is relevant. 
In this case, however, the initial temperature is non-uniform and its jump satisfies Rankine-Hugoniot. 
As a result, the density profile does not relax to a hyperbolic tangent, but rather the incoming flow experiences an overcompression that is established in a characteristic time of a fraction of $\mathcal{T}_{\text{thin}}$, see panel \ref{fig:OptThinAnalysis}(a).
A strong radiant flux is established, which quickly carries away the internal energy gained during compression. 
Consequently, the temperature profile becomes more diffused, leading to the shock formation in quasi-isothermal conditions, as shown by the dashed blue lines.
The resulting shock structure is displaced downstream as the isothermal speed of sound becomes the relevant speed during this time. 
The local bump observed in the temperature profile confirms again second-order corrections in this regime.
The same shock conditions are simulated in panel \ref{fig:OptThinAnalysis}(b) but with larger characteristic lengths and times specific to the optically thick regime. 
In this case we do not capture the opt. thin dynamics, since the disparity in scales between both regimes is in the order of Bo$^{-2} \sim 10^4$. 
Instead, a slight  overcompression is observed, with density eventually relaxing to a hyperbolic tangent shape. 
Matter and radiation temperatures remain in equilibrium at this time scale.
These panels corroborate that in a strongly radiant plasma, radiation can mediate weak shock formation via two distinct regimes, each with disparate formation times. 

\begin{figure}[H]
    \centering
    \includegraphics[width=0.49\textwidth]{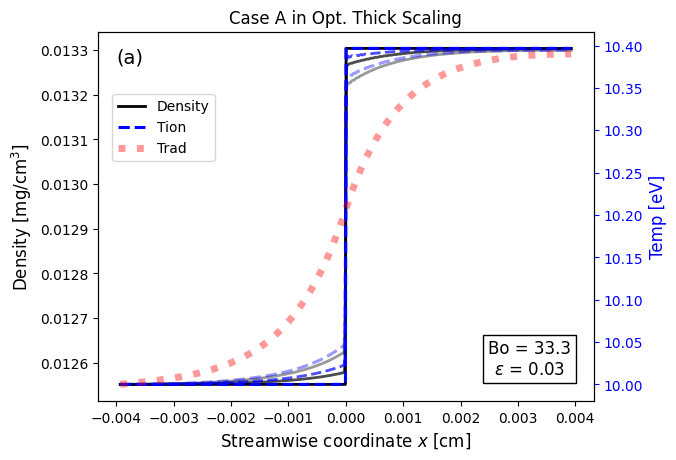}
    \hfill
    \includegraphics[width=0.49\textwidth]{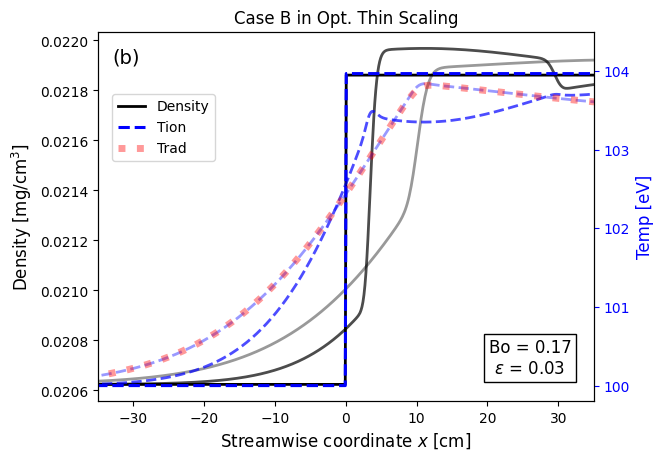}
    \caption{Evolution of density and temperatures from  \textit{FLASH} simulations. (a) Conduction dominated Case A run with a characteristic time and length specific to the optically thick regime. (b) Optically thick, radiation dominated Case B run with a characteristic time and length provided by the optically thin regime. Darker to lighter curves depict increasing simulation time: (a) $t/t_c = 0, 1 $ and 5. (b)  $t/t_c = 0, 0.2$ and 0.6. The normalizing time $t_c$ is different in each panel and given in Table \ref{tab:CaseDescription}. Radiation temperature at the latest plotting time is shown in dashed red lines.}
    \label{fig:CasesDifferentScaling}
\end{figure}

We finally demonstrate in Fig. \ref{fig:CasesDifferentScaling} Cases A and B once more, but this time with characteristic scales that do not correspond appropriately to their respective dissipative mechanisms. 
Fig. \ref{fig:CasesDifferentScaling}(a) proves that radiation cannot support shocks if the Boltzmann number is not small enough.
While a radiatively-driven precursor forms in a time scale much longer than $\mathcal{T}_{\text{cond.}}$, the hydrodynamic shock remains embedded within and mediated by thermal conduction.
In Fig. \ref{fig:CasesDifferentScaling}(b), we inspect the shock formation of a radiative shock (Case B) in a shorter time scale, appropriate to the opt. thin regime. 
Contrary to Fig. \ref{fig:OptThinAnalysis}(a), the incoming flow in this case does not undergo permanent overcompression. 
The radiant flux is not strong enough to dissipate the gained internal energy, and radiation and matter remain in thermal equilibrium. 
We essentially observe a diffusion of profiles across the domain, as the shock develops a thicker structure.
This case confirms that the conditions required for the existence of the optically thin regime are more restrictive than those for the optically thick regime.

This simple setup described provides a validation test to benchmark radiation - hydrodynamic codes. 
\emph{FLASH} has been able to successfully capture the predominant supporting mechanisms and regimes of interaction between matter and radiation when scanning the parameter space $\lbrace \epsilon\ll 1 , \text{Bo}\rbrace$.

\section{\label{Sec: conclusions}Conclusions and discussion}

In this work we have discussed the nature of weak shocks developing in radiative plasmas.
We aim at understanding the formation process of such waves after two plasmas with different pressures are put in contact. 
The smallness of the shock strength has been exploited to derive a set of Burgers-type equations governing the temporal evolution of the perturbed variables. 

We have formally obtained three different solutions corresponding to conduction dominated shocks, radiation dominated optically thick shocks, and radiation dominated optically thin shocks. 
They all develop a tanh-like profile asymptotically in time given by Eqs. \eqref{eq:SolutionDensDim}, \eqref{eq:SolutionDensDimThick} and \eqref{eq:SolutionDensDimThin}, respectively, but present different characteristic lengths and formation times. 
In the first two solutions, matter is compressed quasi-isentropically and the jump in the fluid variables satisfy the Rankine-Hugoniot conditions. 
Contrary to this, in radiation-dominated, optically thin shocks, the downstream flow radiates away any increase in internal energy, and matter is compressed quasi-isothermally. 

Our results for weak shocks share some similarities with the analysis of monochromatic acoustic waves in a radiating fluid, covered in the classical monograph by \citet{mihalas2013foundations}, in that the propagation velocity of the disturbances varies between the isentropic and isothermal speed of sound. 
However, the formation of weak shocks is a somewhat more complex process as the regimes manifest depending on how the Boltzmann number of the medium Bo scales with the shock strength $\epsilon \ll 1$, as shown in Fig. \ref{fig:Chart2}. 
In a weakly radiant medium, characterized by $\text{Bo}\gg \epsilon ^{-1/2}$, the shock structure is given by thermal conduction and is established in a characteristic time $\mathcal{T}_{\text{cond.}}$.
In a medium where $\text{Bo}\lesssim \epsilon ^{-1/2}$, radiation effects are strong enough to modify the conduction-mediated shock in a larger time scale $\mathcal{T}_{\text{thick}}$.
In such media, shocks always develop a wider, optically thick shock structure asymptotically in time.
In strongly radiant media, $\text{Bo}\lesssim \epsilon ^{3/2}$, the shock formation undergoes a transitory regime in which the shock is supported by radiation but develops an optically thin structure. 
Such structure is established in a quicker time scale $\mathcal{T}_{\text{thin}}$ after which the optically thick structure is established. 

This theoretical analysis provides a useful test to benchmark multi-physics codes. 
In this work, we retrieved these asymptotic regimes and scaling laws using the radiation hydrodynamics code \emph{FLASH}. 
The existence of the intermediate radiation dominated, optically thin regime manifests as an overcompression of the material that is observed to occur when $\text{Bo}\lesssim \epsilon ^{3/2}$ in the simulation setup, in agreement with the theoretical predictions. 


Finally, we discuss how the present work can be used to analyse the dynamics of radiative Z-pinch implosions. 
We first notice that the enhancement of a pinch compression due to radiative losses is the core of an extensively discussed phenomenon known as radiative collapse.\cite{haines1989analytic, velikovich1995implosions}
In its theoretical derivations, a single optically thin Z-pinch column is generally considered under the assumption of time-dependent pressure equilibrium. 
The present work focuses rather on the nature of compressional waves encompassing arbitrary optical depths, and as such is better suited to inspect the liner physics in liner-on-target configurations. 

\begin{figure}[H]
    \centering
         \includegraphics[width=0.6\textwidth]{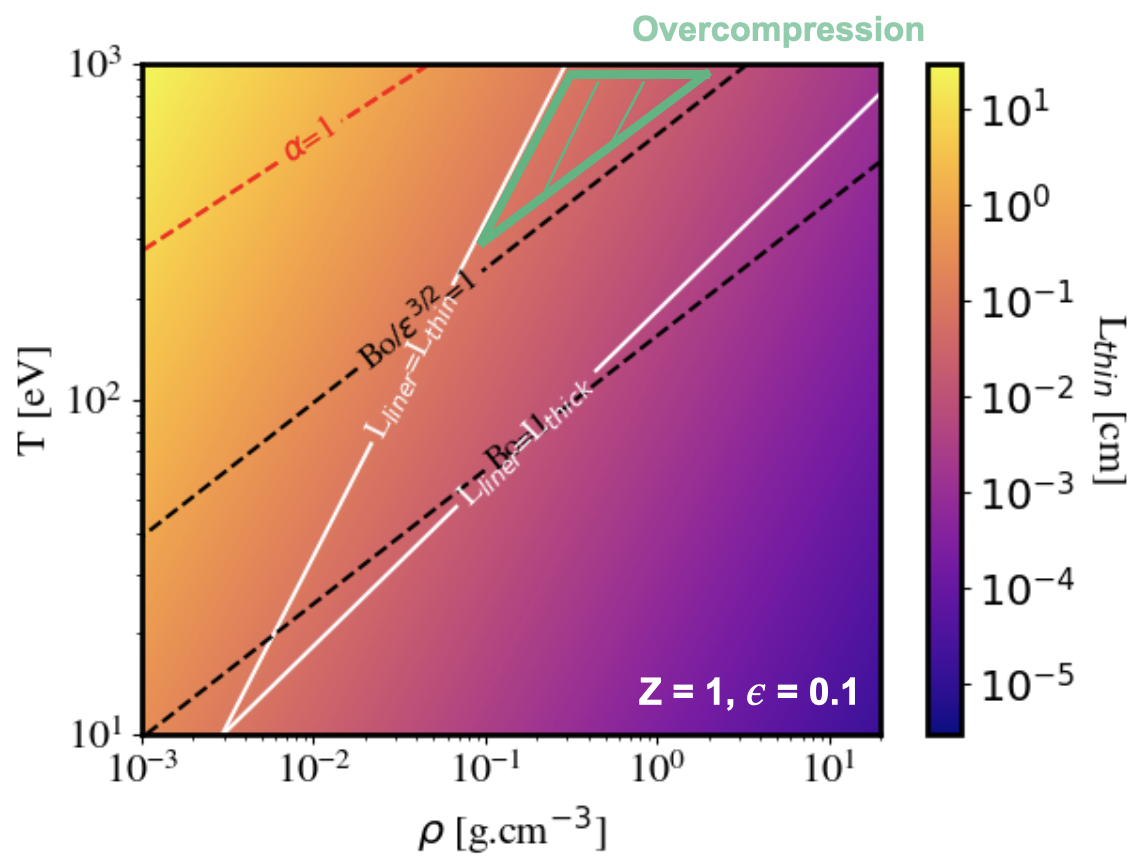}
         \includegraphics[width=0.6\textwidth]{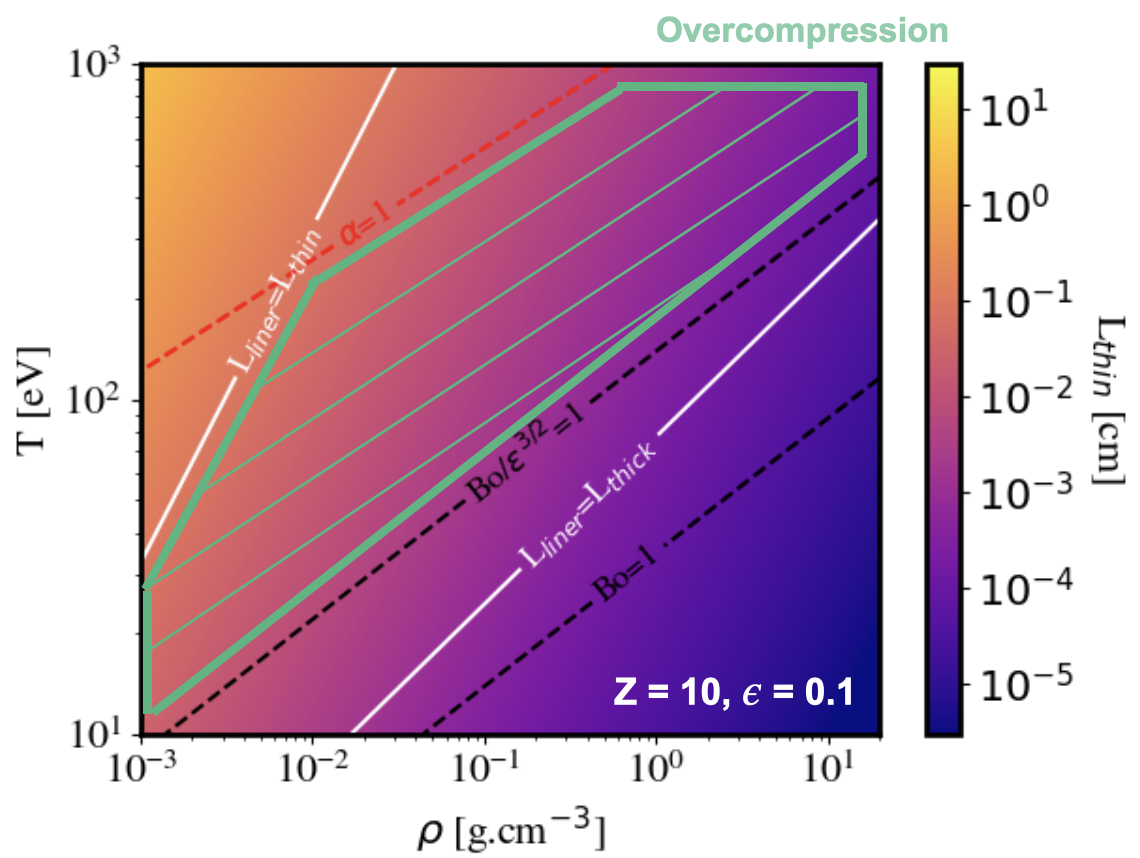}
    \caption{Isocontours of relevant parameters and lengths in the design space of liner density and temperature. A liner thickness $L_{\text{liner}} = 0.1$ cm and a shock strength $\epsilon = 0.1$ are considered. Top panel: Deuterium liner Z = 1. Bottom panel: Xenon liner with an ionization state Z = 10. The green shadowed region corresponds to the conditions in which such shock would develop a radiation dominated, optically thin structure, leading to overcompression. The red dotted line $\alpha = 1$ corresponds to internal pressure equal to radiation pressure. }
    \label{fig:Chart4ab}
\end{figure}

Figure \ref{fig:Chart4ab} illustrates in which region of the parameter space defined by the liner density and temperature are compressional waves radiation-dominated and optically thin.
Throughout the chart, we have considered a constant ionization state but retained the dependence on temperature and density in the photon mean-free-path.
We take as an example a pressure wave of strength $\epsilon = 0.1$ launched in a liner of thickness $L_{\text{liner}} = 0.1$ cm. 
This work shows that such a wave would lead to overcompression according to following considerations.
The liner thickness should be larger than $\mathcal{L}_{\text{thin}}$ to ensure that radiation can effectively support this weak shock. 
However, the liner thickness should not exceed $\mathcal{L}_{\text{thick}}$ or the radiant flux emitted by the wave will entirely be absorbed within the liner, preventing overcompression.
The white lines in Fig. \ref{fig:Chart4ab} delineate these two conditions.
Additionally, the characteristic Boltzmann number of the liner should satisfy $\text{Bo}\lesssim \epsilon ^{3/2}$ for the optically thin regime to exist, which sets the lower bound to the dashed black line.
Finally, the radiation pressure should remain negligible to be consistent with the hypothesis of this analysis and to avoid employing driver energy in compressing the radiation field.
A small radiation-to-internal pressure ratio $\alpha$ is ensured below the dashed red line. 

The region delimited by such considerations is shadowed in green in Fig. \ref{fig:Chart4ab} both for a deuterium and a xenon gas-puff liner.
In the latter, a ionization state $Z=10$ is considered. 
Despite the simplifying assumptions of equal Rosseland and Planck mean free paths, being both of them given by bremsstrahlung physics solely (which is specially inaccurate in the xenon case), this chart showcases the benefit of using high-atomic-number liners to enhance compression.
Of course, other considerations such as liner stability or fuel-liner mixing should also be present in the liner material choice.  
In any case, this analysis provides a first estimation to assess conditions that enable liner adiabat control through radiative effects.

\begin{acknowledgments}
This material is based upon work supported by the Department of Energy (DOE) National Nuclear Security Administration (NNSA) under Award Numbers DE-NA0003856, DE-NA0003842, DE-NA0004144, and DE-NA0004147, under subcontracts no. 536203 and 630138 with Los Alamos National Laboratory, and under subcontract B632670 with Lawrence Livermore National Laboratory. We acknowledge support from the U.S. DOE Advanced Research Projects Agency-Energy (ARPA-E) under Award Number DE-AR0001272 and the U.S. DOE Office of Science under Award Number DE-SC0023246. The software used in this work was developed in part by the U.S. DOE NNSA- and U.S. DOE Office of Science-supported Flash Center for Computational Science at the University of Chicago and the University of Rochester.
\end{acknowledgments}


\bibliography{mylibrary}

\end{document}